\documentstyle[preprint,aps]{revtex}

\begin{document}

\preprint{SLAC-PUB-6431}

\title{Wavefunction-Independent
Relations between the Nucleon Axial-Coupling
$g_A$ and the Nucleon Magnetic Moments}
\author{Stanley J. Brodsky and Felix Schlumpf}
\address{
Stanford Linear Accelerator Center\\
Stanford University, Stanford, California 94309
}
\date{\today}
\maketitle

\begin{abstract}
We calculate the proton's magnetic moment $\mu_p$ and
its axial-vector coupling
$g_A$  as a function of its
Dirac radius $R_1$ using a relativisitic
three-quark model formulated on the light-cone. The relationship
between $\mu_p$ and $g_A$ is found to be
independent of the assumed form of the light-cone wavefunction.
At the physical radius $R_1=0.76$ fm, one obtains the experimental
values for both $\mu_p$ and $g_A$,
and the helicity carried by the valence $u$
and $d$ quarks are each reduced by a
factor  $\simeq 0.75$ relative to their
non-relativistic values.
At large
proton radius, $\mu_p$ and $g_A$ are given by the usual non-relativistic
formulae. At small radius, $\mu_p$ becomes equal to the Dirac
moment, as demanded by the
Drell-Hearn-Gerasimov sum rule. In addition, as $R_1
\to 0,$ the  constituent
quark helicities become completely disoriented and
$g_A \to 0$.

\end{abstract}


\narrowtext

In recent years the light-cone quantization of quantum-field theory has
emerged as a promising method for solving
relativistic bound-state problems in
the strong coupling regime~\cite{brod93}.
Light-cone quantization has a number of
unique features that make it appealing,
most notably, the ground state of the
free theory is also a ground state of the
full theory, and the Fock expansion
constructed on this vacuum state provides a complete relativistic
many-particle basis for diagonalizing the full theory. The method seems
therefore to be well-suited to solving quantum chromodynamics. For
practical calculations one approximates  the field theory by truncating
the Fock space~\cite{perry}. The assumption is that a few excitations
describe the
essential physics, and that adding more excitations only refines this
initial approximation.
This is quite different from the instant formulation
of QCD where an infinite
number of gluons is essential for formulating even
the vacuum. In this paper
we restrict ourselves to an effective three-quark Fock
description of the nucleon.
In this effective theory, all additional degrees of
freedom (including zero modes) are parameterized in an effective
potential~\cite{lepa80}.
In such a theory the constituent quarks will also
acquire effective masses and form factors.

After  truncation, one could in principle obtain the mass $M$ and
light-cone wavefunction
$|\Psi \rangle$
of the
three-quark bound-states by solving the
Hamiltonian eigenvalue problem
\begin{equation}
H^{\rm effective}_{\rm LC} |\Psi \rangle
= M^2 |\Psi \rangle. 
\end{equation}
Given  the eigensolutions
$|\Psi \rangle$ one could then compute the form factors and
other properties of the baryons.
Even without explicit solutions, one knows that
the helicity and flavor structure of the
baryon eigenfunctions must reflect the assumed
global SU(6) symmetry and Lorentz
invariance of the theory. However, since we do not
have an explicit representation
for the effective potential in the light-cone
Hamiltonian $H^{\rm effective}_{\rm LC}$ for three-quarks,
we shall have to proceed by
making an ansatz for the momentum
space structure of the wavefunction $\Psi.$
This may seem quite arbitrary,
but as we will show below, for a given size of
the proton, the predictions and
interrelations between observables at $Q^2=0,$
such as the proton magnetic moment $\mu_p$
and its axial coupling $g_A,$ turn
out to be essentially independent of the shape of the wavefunction.

The light-cone model given in
Ref.~\cite{schl93} provides a framework for representing the general
structure of the effective  three-quark wavefunctions for baryons.
The wavefunction
$\Psi$ is constructed as the product of a
momentum wavefunction, which is
spherically symmetric and invariant under permutations,
and a spin-isospin wave
function, which is uniquely determined by SU(6)-symmetry requirements.  A
Wigner~\cite{wigner} (Melosh~\cite{melosh}) rotation
is applied to the spinors,
so that the wavefunction of the proton is an eigenfunction
of $J$ and $J_z$ in
its rest frame~\cite{coester,stern}.  To represent the
range of uncertainty in
the possible form of the momentum wavefunction we
choose two simple functions of
the invariant mass ${\cal M}$ of the quarks:
\begin{eqnarray} \psi_{\rm H.O.}({\cal
M}^2)&=&N_{\rm H.O.}\exp(-{\cal M}^2/2\beta^2),\\ \psi_{\rm Power}({\cal
M}^2)&=&N_{\rm Power}(1+{\cal M}^2/\beta^2)^{-p} \end{eqnarray}
where $\beta$ sets the scale of the nucleon size.
 Perturbative QCD predicts
a nominal power-law fall off at large $k_\perp$ corresponding to
$p=3.5$~\cite{lepa80}. The invariant mass ${\cal M}$ can be written as
\begin{equation}
{\cal M}^2 = \sum_{i=1}^3 \frac{\vec k_{\perp i}^2+m^2}{x_i}
\end{equation}
where we used the longitudinal light-cone
momentum fractions $x_i=p_i^+/P^+$ ($P$
and $p_i$ are the nucleon and quark momenta,
respectively, with $P^+=P_0+P_z$).
The internal momentum variables $\vec k_{\perp i}$ are given by
$\vec k_{\perp
i}=\vec p_{\perp i}-x_i \vec P_\perp$ with the
constraints $\sum \vec k_{\perp i}=0$
and $\sum x_i=1$.  The Melosh rotation has the matrix
representation~\cite{melosh}
\begin{equation}
R_M(x_i,k_{\perp i},m)=\frac{m+x_i {\cal M}-i\vec \sigma\cdot(\vec
n\times \vec  k_i)}
{\sqrt{(m+x_i {\cal M})^2+\vec k_{\perp i}^2} },
\end{equation}
with $\vec  n=(0,0,1)$, and it becomes the unit matrix if
the quarks are collinear
\begin{equation}
R_M(x_i,0,m)=1 .
\end{equation}
Thus the internal transverse momentum dependence of the light-cone
wavefunctions also affects its helicity structure.

The Dirac and Pauli form factors
$F_1(Q^2)$ and $F_2(Q^2)$ of the nucleons are
given by the spin-conserving and the spin-flip vector current
$J^+_V$ matrix elements ($Q^2=-q^2$) \cite{brod80}
\begin{eqnarray}
F_1(Q^2) &=& \langle p+q,\uparrow | J^+_V | p,\uparrow \rangle , \\
(Q_1-i Q_2) F_2(Q^2) &=& -2M\langle p+q,\uparrow | J^+_V | p,
\downarrow \rangle .
\end{eqnarray}
We then can calculate the anomalous magnetic moment $a=\lim_{Q^2\to 0}
F_2(Q^2)$.  [The total proton magnetic
moment is  $\mu_p = {e \over 2M}(1+a_p).$]
The same parameters as in Ref.~\cite{schl93} are chosen;
namely $m=0.263$ GeV (0.26 GeV) for the up- and down-quark masses, and
$\beta=0.607$ GeV (0.55 GeV)
for $\psi_{\rm Power}$ ($\psi_{\rm H.O.}$) and
$p=3.5$.
The quark currents are taken as elementary currents with Dirac moments
${e_q \over 2 m_q}.$
All of the baryon moments are
well-fit if one takes the strange quark mass as 0.38 GeV.
With the above values, the proton magnetic moment
is 2.81 nuclear magnetons,
the neutron magnetic moment is $-1.66$ nuclear
magnetons\footnote{The neutron value can be improved by
relaxing the assumption of isospin symmetry.} and
the radius of the proton is 0.76 fm; i.e., $M_p R_1=3.63$ \cite{schl93}.

In Figure~1 we show the functional relationship between the
anomalous moment $a_p$ and its Dirac
radius predicted by the three-quark light-cone model.
The value of
$R^2_1 = -6 dF_1(Q^2)/dQ^2\vert_{Q^2=0}$ is varied by
changing $\beta$ in the
light-cone wavefunction while keeping the quark mass
$m$ fixed.   The prediction for
the power-law wavefunction
$\psi_{\rm Power}$ is given by the broken line;
the continuous line represents
$\psi_{\rm H.O.}$.
Figure~1 shows that when one plots the dimensionless
observable $a_p$  against
the dimensionless observable $M R_1$ the prediction is essentially
independent of the assumed power-law or Gaussian form of the three-quark
light-cone wavefunction.  Different values
of $p > 2 $ do also not affect
the functional dependence of $a_p(M_p R_1)$ shown in Fig.~1.
In this sense the
predictions of the  three-quark  light-cone model  relating the
$Q^2 \to 0$ observables are essentially model-independent.
The only parameter controlling the relation between the dimensionless
observables in the light-cone three-quark model is $m/M_p$
which is set to 0.28.
For the physical
proton radius $M_p R_1=3.63$
one obtains the empirical value for $a_p=1.79$
(indicated by the dotted lines in Figure~1).

The prediction for the anomalous moment
$a$ can be written analytically  as
$a=\langle \gamma_V \rangle a^{\rm NR}$,
where $a^{\rm NR}=2M_p/3m$ is the
non-relativistic ($R\to\infty$) value and $\gamma_V$ is
given as~\cite{chung}
\begin{equation}
\gamma_V(x_i,k_{\perp i},m)=
\frac{3m}{{\cal M}}\left[ \frac{(1-x_3){\cal M}(m+x_3 {\cal M})-
\vec k_{\perp 3}^2/2}{(m+x_3 {\cal M})^2+\vec k_{\perp 3}^2} \right] .
\end{equation}
The expectation  value $\langle \gamma_V \rangle$ is evaluated
as\footnote{
$[d^3k]=d\vec k_1d\vec k_2d\vec k_3\delta(\vec k_1+ \vec k_2+ \vec k_3)$.
The third component of $\vec k$ is defined as $k_{3i}=
\frac{1}{2}(x_i{\cal M}-\frac{m^2+\vec k_{\perp i}^2}{x_i {\cal M}})$.
This measure differs from the usual one used in Ref.~\protect\cite{lepa80}
by the Jacobian $\prod \frac{dk_{3i}}{dx_i}$ which can be
absorbed into the wavefunction.}
\begin{equation}
\langle \gamma_V \rangle =
\frac{\int [d^3k] \gamma_V |\psi|^2}{\int [d^3k] |\psi|^2} .
\end{equation}

We now take a closer look at the two limits $R \to \infty$ and $R\to 0$.
In the non-relativistic limit we let $\beta \to 0$ and keep the quark mass
$m$ and the proton mass $M_p$ fixed.
In this limit the proton radius $R_1 \to \infty$ and $a_p \to
2M_p/3m = 2.38$ since $\langle \gamma_V \rangle \to 1$\footnote{This
differs slightly from the usual non-relativistic formula $1+a=\sum_q
\frac{e_q}{e} \frac{M_p}{m_q}$ due to the non-vanishing binding
energy which results in $M_p \neq 3m_q$.}.
Thus the physical value of the anomalous magnetic moment at the
empirical proton radius
$M_p R_1=3.63$
is reduced
by 25\% from its
non-relativistic value due to relativistic recoil and nonzero
$k_\perp$\footnote{The non-relativistic value of the neutron
magnetic moment is reduced by 31\%.}.

To obtain the
ultra-relativistic limit we let $\beta \to \infty$ while keeping
$m$ fixed.  In this limit the proton becomes pointlike $M_p R_1 \to 0$
and the
internal transverse momenta $k_\perp \to \infty$.
The anomalous magnetic
momentum of the proton goes linearly to zero as $a=0.43 M_p R_1$ since
$\langle \gamma_V \rangle \to 0$.  Indeed, the
Drell-Hearn-Gerasimov (DHG) sum
rule \cite{dhg} demands that the proton magnetic moment becomes equal to
the Dirac moment at small radius.  For a spin-$1\over 2$ system
\begin{equation}
a^2=\frac{M^2}{2\pi^2\alpha}\int_{s_{th}}^\infty
\frac{ds}{s}\left[ \sigma_P(s)-\sigma_A(s)\right],
\label{eq:dhg}
\end{equation}
where $\sigma_{P(A)}$ is the total photoabsorption cross section with
parallel (antiparallel) photon and target spins.
If we take the point-like
limit, such that the threshold for inelastic excitation becomes infinite
while the mass of the system is kept finite, the integral over the
photoabsorption cross section vanishes and $a=0$ \cite{brod80}.  In
contrast, the anomalous magnetic moment of the proton does not vanish
in the non-relativistic quark model as $R\to 0$. The
non-relativistic quark model does not reflect the fact that the magnetic
moment of a baryon is derived from lepton scattering at
non-zero momentum transfer \cite{primack}; i.e.,
the calculation of a magnetic
moment requires knowledge of the boosted wavefunction.
The Melosh transformation
is also essential for deriving the DHG sum rule and
low energy theorems (LET) of
composite systems~\cite{primack}.

A similar analysis can be performed for the
axial-vector coupling measured in neutron decay.
The coupling $g_A$ is given
by the spin-conserving axial current $J_A^+$ matrix element
\begin{equation}
g_A(0) = \langle p,\uparrow | J^+_A | p,\uparrow \rangle.
\end{equation}
The value for $g_A$ can be written
as $g_A=\langle \gamma_A \rangle g_A^{\rm
NR}$ with $g_A^{\rm NR}$ being the non-relativistic value of $g_A$ and
with $\gamma_A$ as~\cite{chung,ma}
\begin{equation}
\gamma_A(x_i,k_{\perp i},m)=\frac{(m+x_3 {\cal M})^2-\vec k_{\perp 3}^2}
{(m+x_3 {\cal M})^2+\vec k_{\perp 3}^2} .
\label{eq:ma}
\end{equation}
In Fig.~2 the axial-vector coupling is plotted against the proton radius
$M_p R_1$.  The same parameters and the same line
representation as in Fig.~1
are used.  The functional dependence of $g_A(M_p R_1)$
is also found to be independent
of the assumed wavefunction.
At the physical proton radius $M_p R_1=3.63$ one
predicts the value
$g_A = 1.25$ (indicated by the dotted lines in
Figure~2) since $\langle \gamma_A
\rangle =0.75$.   The measured value is
$g_A= 1.2573\pm 0.0028$~\cite{PDG}.
This is a 25\%
reduction compared to the non-relativistic SU(6)  value $g_A=5/3,$
which is only valid for a proton with large radius $R_1 >> 1/M_p.$
As shown by Ma and Zhang~\cite{ma} the Melosh rotation
generated by the internal transverse momentum
spoils the usual
identification of the $\gamma^+ \gamma_5$ quark current matrix
element with the total rest-frame spin projection $s_z$, thus resulting
in a reduction of $g_A$.

Thus given
the empirical
values for  the proton's anomalous moment $a_p$
and radius $M_p R_1,$
its axial-vector coupling is automatically fixed at the value
$g_A=1.25.$   This
prediction is an essentially
model-independent  prediction of the three-quark
structure of the proton in QCD. The Melosh rotation of the light-cone
wavefunction is crucial for
reducing the value of the axial coupling from its
non--relativistic value 5/3 to its empirical value.
In Figure~3 we plot
$g_A/g_A(R_1 \to \infty)$ versus $a_p/a_p(R_1 \to \infty)$ by varying the
proton radius $R_1.$ The near equality of
these ratios reflects the
relativistic spinor structure of the nucleon bound state,
which is essentially independent of the detailed shape of the
momentum-space
dependence of the light-cone
wavefunction.

We emphasize
that at small proton radius the light-cone model predicts
not only a vanishing anomalous moment but also
\begin{equation}
\lim_{R_1 \to 0} g_A(M_p R_1)=0.
\end{equation}
One can understand this physically: in the zero radius limit the internal
transverse momenta become infinite and the quark helicities
become completely
disoriented.  This is in contradiction with chiral models
which suggest that
for a zero radius composite baryon one should obtain the
chiral symmetry result
$g_A=1$.

The helicity measures $\Delta u$ and $\Delta d$ of the
nucleon each experience the same reduction as $g_A$
due to the Melosh effect.
Indeed, the quantity $\Delta q$ is defined by the axial
current matrix element
\begin{equation}
\Delta q=\langle p,\uparrow | \bar q\gamma^+\gamma_5 q | p,\uparrow
\rangle ,
\label{eq:dq}
\end{equation}
and the value for $\Delta q$ can be written analytically as $\Delta
q=\langle \gamma_A \rangle \Delta q^{\rm NR}$ with $\Delta q^{\rm NR}$
being the non-relativistic or naive value of $\Delta q$ and with
$\gamma_A$ given in Eq.~(\ref{eq:ma}).

Figure~4 shows the prediction of the
light-cone model for the quark helicity sum
$\Delta\Sigma=\Delta u+\Delta d$ as a
function of the proton radius $R_1$.
The same parameters and the same line
representation as in Fig.~1 are used.
This figure shows that the helicity sum $\Delta\Sigma$ defined from the
light-cone wavefunction  depends on the proton size,
and thus it cannot be
identified as the vector sum of the rest-frame constituent spins. As
emphasized by Ma~\cite{ma}, the rest-frame spin sum is not a Lorentz
invariant  for a composite system.
Empirically, one can measures $\Delta q$ from
the first moment of the leading twist polarized structure function
$g_1(x,Q).$  In the light-cone and parton model
descriptions, $\Delta q=\int_0^1
dx [q^\uparrow (x) - q^\downarrow (x)]$, where $q^\uparrow (x)$ and
$q^\downarrow (x)$ can be interpreted as the probability for finding a
quark or antiquark  with longitudinal momentum
fraction $x$  and polarization
parallel or antiparallel to the proton helicity
in the proton's infinite momentum
frame~\cite{lepa80}.
[In the infinite momentum there is no distinction between
the quark helicity and its spin-projection $s_z.$]
Thus $\Delta q$ refers to the
difference of  helicities at fixed light-cone time
or at infinite momentum; it
cannot be  identified with
$q(s_z=+\frac{1} {2})-q(s_z=-\frac{1}{2}),$ the spin
carried by each quark flavor in the proton rest frame in the equal time
formalism.

One sees from figure~4 that the usual SU(6) values $\Delta u^{\rm
NR}=4/3$ and $\Delta d^{\rm NR}=-1/3$ are only valid predictions for
the proton at large $M R_1.$
At the physical radius the quark helicities are
reduced by the same ratio 0.75 as $g_A/g_A^{\rm NR}$ due to the Melosh
rotation. Qualitative arguments for such a reduction have been given in
Refs.~\cite{karl,fritzsch}.
 Thus for  $M_p R_1 = 3.63,$ the
three-quark model predicts $\Delta u=1,$ $\Delta d=-1/4,$ and
$\Delta\Sigma=\Delta u+\Delta d  = 0.75$.
Although the gluon contribution
$\Delta G=0$ in our model, the general sum rule~\cite{jaffe}
\begin{equation}
\frac{1}{2}\Delta \Sigma +\Delta G+L_z= \frac{1}{2} 
\end{equation}
is still satisfied,
since the Melosh transformation effectively contributes to
$L_z$.

Suppose one adds polarized gluons to the three-quark light-cone  model.
Then the flavor-singlet quark-loop radiative
corrections to the gluon propagator
will give an  anomalous contribution $\delta
(\Delta q)=-\frac{\alpha_s}{2\pi}\Delta G$ to each light quark
helicity~\cite{Altarelli}.  The predicted
value of $g_A = \Delta u - \Delta d$  is of course unchanged.
For illustration
we shall choose $\frac{\alpha_s}{2\pi}\Delta G=0.20$.
The gluon-enhanced quark
model then gives the values in Table~1,
which agree well with the present
experimental values. Note that the gluon anomaly
contribution to $\Delta s$ has
probably been overestimated here due to the large strange quark mass.
One could also envision other sources for this shift of $\Delta q$
such as intrinsic flavor~\cite{fritzsch}.

In summary, we have shown that relativistic effects are important for
understanding the spin structure of the nucleons.
By plotting dimensionless observables
against dimensionless observables
we obtain model-independent relations independent of the
momentum-space form of the three-quark light-cone wavefunctions.
For example,
the value of $g_A \simeq 1.25$ is correctly predicted from the
empirical value
of the proton's anomalous moment. For the physical
proton radius $M_p R_1= 3.63$ the
inclusion of the Wigner (Melosh) rotation due to  the finite  relative
transverse momenta of the three quarks results
in a  $\simeq 25\% $
reduction of the non-relativistic predictions for the anomalous magnetic
moment, the axial vector coupling, and the quark helicity content of the
proton.

\acknowledgments

This work was supported in part by the Schweizerischer Nationalfonds and
in part by the Department of Energy, contract DE-AC03-76SF00515. We thank
Michael Boulware, Harald Fritzsch,
Michael Peskin, and Ivan Schmidt for helpful discussions.

\narrowtext
\begin{table}
\caption{Comparison of the quark content of the proton in the
non-relativistic quark model (NR), in our three-quark model (3q), in
a gluon-enhanced three-quark  model (3q+g), and with experiment
\protect\cite{ek}.}
\begin{tabular}{crrrc}
Quantity&NR& 3q & 3q+g & Expt. \\
\tableline
$\Delta u$ & $\frac{4}{3}$ & 1 & 0.80 & $0.80\pm 0.04 $ \\
$\Delta d$ &$-\frac{1}{3}$ & $-\frac{1}{4}$ & --0.45& $-0.46\pm 0.04 $ \\
$\Delta s$ & 0 & 0 & --0.20 & $-0.13\pm 0.04 $ \\
$\Delta \Sigma$ &1 &  $\frac{3}{4}$ & 0.15 & $0.22\pm 0.10 $ \\
\end{tabular}
\label{table:1}
\end{table}

\begin{figure}
\caption{The anomalous magnetic moment $a=F_2(0)$ of the proton as a
function of $M_p R_1$: broken line, pole type wavefunction; continuous
line, gaussian wavefunction. The experimental value is given by the
dotted lines. Our model is independent of the wavefunction for $Q^2=0$.
\label{fig:f2}}
\end{figure}

\begin{figure}
\caption{The axial vector coupling $g_A$ of the neutron to proton decay as
a function of $M_p R_1$: line code as in Fig.~1.  The experimental value is
given by the dotted lines.
\label{fig:ga}}
\end{figure}

\begin{figure}
\caption{$g_A/g_A(R_1 \to \infty)$ versus $a_p/a_p(R_1 \to \infty)$
by varying the proton radius $R_1.$: line code as in Fig.~1.
\label{fig:}}
\end{figure}

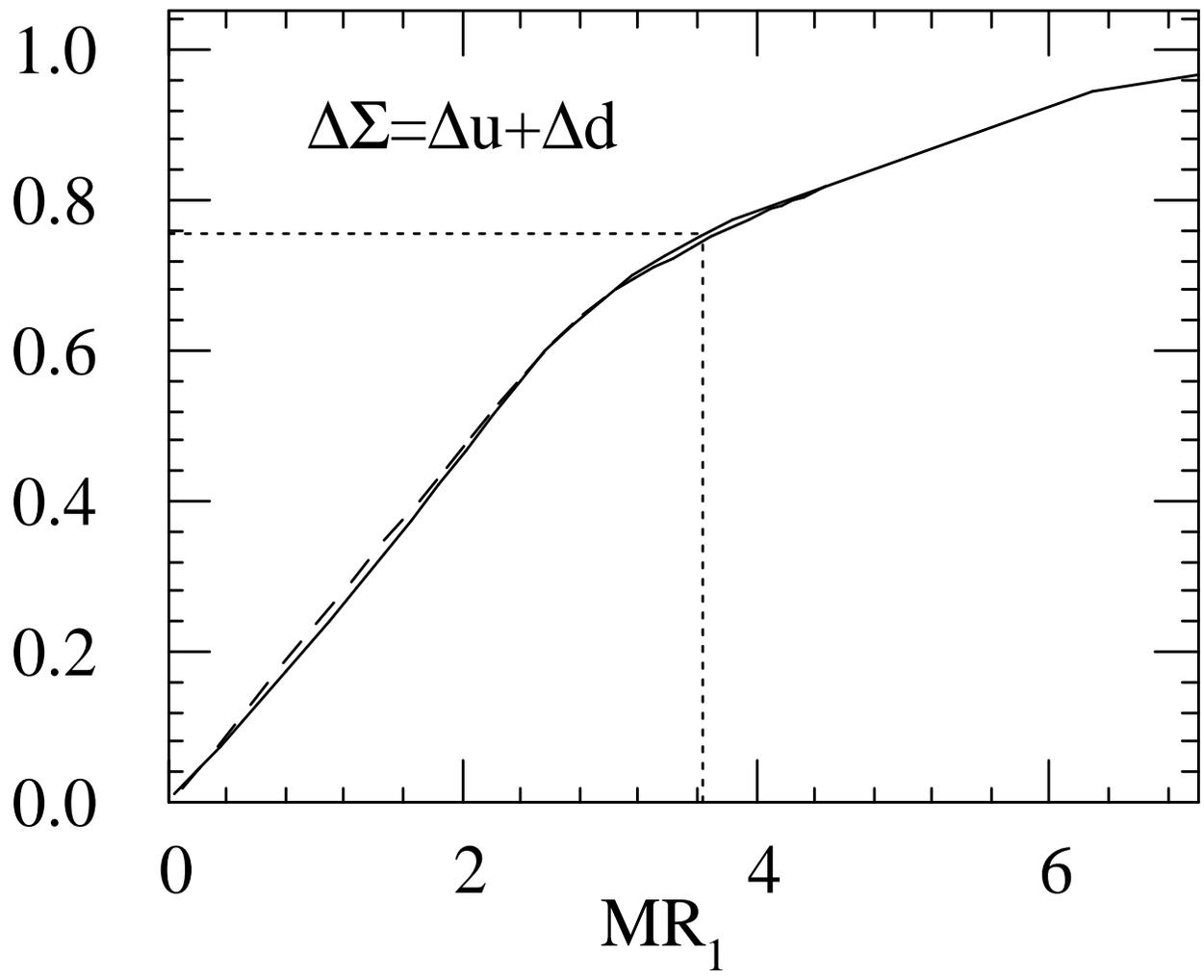
\begin{figure}
\caption{The quantity $\Delta\Sigma=\Delta u+\Delta d$ of the proton as
a function of $M_p R_1$: line code as in Fig.~1.
\label{fig:dq}}
\end{figure}

\end{document}